Designing a single-molecule biophysics tool for characterizing DNA damage for techniques that kill infectious pathogens through DNA damage effects

Helen Miller, Adam J. M. Wollman, Mark C. Leake

Definitions:

Intercalating = binding between base pairs

PCR= Polymerase chain reaction; a method to produce multiple copies of a DNA sequence. The DNA to be copied is heated to separate the strands, and then cooled so that short 'primer' sequences bind complementarily at both 5' ends of the DNA sequence to be copied. A DNA polymerase then extends from the primers to give two copies of the sequence of interest. Multiple repeats of this process produce a high yield of the target sequence.

Abstract

Antibiotics such as the quinolones and fluoroquinolones kill bacterial pathogens ultimately through DNA damage. They target the essential type IIA topoisomerases in bacteria by stabilising the normally transient double strand break state which is created to modify the supercoiling state of the DNA. Here we discuss the development of these antibiotics and their method of action. Existing methods for DNA damage visualisation, such as the comet assay and immunofluorescence imaging can often only be analysed qualitatively and this analysis is subjective. We describe a putative single-molecule fluorescence technique for quantifying DNA damage via the total fluorescence intensity of a DNA origami tile fully saturated with an intercalating dye, along with the optical requirements for how to implement these into a light microscopy imaging system capable of single-molecule millisecond timescale imaging. This system promises significant improvements in reproducibility of the quantification of DNA damage over traditional techniques.

Introduction

There exist classes of antibiotics that act to kill bacterial pathogens via DNA damage. These antibiotics, and existing methods to measure DNA damage will be discussed, along with considerations for the design of a novel single-molecule fluorescence method of measuring DNA damage.

1 Quinolones: Antibiotics that kill pathogens via DNA damage

The quinolones, and more recently fluoroquinolones, are synthetic antibiotics which kill infectious pathogens through DNA damage by stabilising double strand breaks (DSB) caused by the type IIA topoisomerases. The first quinolone discovered was Nalidixic Acid (Lesher et al. 1962). The quinolone antibiotics are effective against gram-negative bacteria, being bacteriostatic (stopping growth) at low concentrations and bactericidal at higher concentrations when hydroxyl radicals are formed (Kohanski et al. 2007) (a detailed discussion of the activities of the different drugs is given in (Drlica and Zhao 1997)). Fluoroquinolones, such as the first one discovered, Norfloxacin, have fluorine attached to the ring system (see figure 1, for details on the chemical structure of various quinolone and fluoroquinolone compounds see (Gootz and Brighty 1996; Appelbaum and Hunter 2000)), and have increased activity in gram-positive bacteria compared to the quinolones (Chu and Fernandes 1991).

Fluoroquinolones have several major advantages over quinolones: they generally are more effective against gram-positive bacteria with the same activity against gram-negative bacteria, and they are absorbed well enough and the newer ones have a sufficiently long half-life that a daily dose is possible which increases patient compliability (Appelbaum and Hunter 2000). However, fluoroquinolones can have toxic side effects (for a review see for Owens & Ambrose, 2005), leading to some being abandoned in the late stages of clinical trials or even after marketing (e.g temafloxacin (Hardy et al. 1987)).

There are two main targets of quinolone antibiotics, both of which are type IIA topoisomerases: DNA gyrase (the target in gram negative bacteria) and topoisomerase IV (the target in gram positive bacteria). DNA gyrase is a tetrameric protein of the form $A_2B_2$ (Sugino et al. 1977) which acts to produce negative supercoils in DNA (Gellert et al. 1976). The A subunits bind to DNA whilst the B subunits carry out the ATP hydrolysis (Reece and Maxwell 1991). Topoisomerase IV is a tetrameric protein of the form $C_2E_2$, which is essential for separating the linked DNA strands produced during chromosome replication (Kato et al. 1990). The C unit binds DNA and the E subunit carries out ATP hydrolysis (Peng and Marians 1993). The C subunit of topoisomerase IV can be thought of as analogous to the A subunit of DNA gyrase, E and B likewise. The main difference in the way these two enzymes act is that DNA gyrase wraps the DNA around itself whilst topoisomerase IV does not (Drlica and Zhao 1997).

Type IIA topoisomerases create transient four base pair staggered DSB through which duplex DNA can be passed to modify supercoiling. A covalent bond is formed between the 5' phosphate of the new DNA terminus and a tyrosine residue in the functional subunit of the type IIA topoisomerase (Sutcliffe et al. 1989; Roca 1995) – to be explicit, in DNA gyrase, the hydroxyl group of Tyrosine-122 and the 5' DNA terminus form a phosphate ester (Horowitz and Wang 1987; Maxwell 1992).

The bacteriostatic method of action of the quinolones is to stabilise the double strand break, putatively by causing a conformational change in the enzyme – DNA complex (Orphanides and Maxwell 1994). Studies with Moxifloxacin, an example of a fluoroquinolone, have shown that it is wedge-shaped molecule that intercalates at the points of DNA breakage; the interaction between the antibiotic and the topoisomerase is mediated by a magnesium ion (Wohlkonig et al. 2010). DNA sedimentation experiments on quinolone treated DNA(Chen et al. 1996) indicate that the ends of the DNA are released, possibly by being freed from the quinolone-topoisomerase complex, or by separation of the subunits of DNA gyrase (Drlica and Zhao 1997), this leaves the bacterial chromosome unwound (at the clinical dose levels used mammalian cells are not affected), leaving it too large to fit in the daughter cells, stopping growth (Emmerson and Jones 2003). The bacterial cell starts the SOS response; an error-prone DNA damage repair system which leads to the fast development of antibiotic resistance. The bactericidal drugs also produce hydroxyl radicals which kill the cell (Kohanski et al. 2007). For more detail on the SOS response, see, for example (Drlica and Zhao 1997; Žgur-Bertok 2013).

2 Quantifying DNA damage

Quantifying the DNA damage required to kill a bacterial pathogen could lead to lower doses of antibiotics being required, which might help with the increasing problem of antibiotic resistance to the quinolones.

The quinolones act by DNA damage, so a method to quantify this potentially gives a way to measure the effectiveness of this class of antibiotics, or to determine the dose required. There are other applications where quantitative measures of DNA damage are required, for example in studying programmed cell death, or for evaluating the effects of novel cancer treatments. For many medical applications it is important to know how different cells respond to a new therapy, by cell type or by location relative to the point of application. This is imperative in developing *in vivo* treatments for cancer, where you only wish to damage the tumour cells and not the healthy tissue cells; you must measure how far from the point of application a cell-killing effect is received – you do not wish to damage healthy tissue. This is important in the application of, for example, medical plasma treatments (Hirst et al. 2015).

2.1 Gel electrophoresis and the comet assay

Two common assays for *in vivo* and *in vitro* quantification of DNA damage work via electrophoresis. DNA is negatively charged and so will migrate through a porous medium when a potential difference is applied. In electrophoresis the distance travelled by a particular fragment depends on both its molecular weight and its conformation. For DNA of the same conformation a strand with lower molecular weight runs further from its starting position than a strand of high molecular weight. DNA of the same molecular weight in supercoiled, circular or linear forms runs to different positions that can be used to identify it's structure, and therefore if it is undamaged, has a single strand break, or a double strand break (Blazek et al. 1989; O'Connell et al. 2011). The DNA in the gels can be visualised either by adding a DNA intercalating dye to the gel before running electrophoresis or by post staining, and imaging with ultraviolet light. As an example of the information that can be gained from electrophoresis figure 2 shows a 1% agarose gel with damage to λ DNA. In Lane 1 the λ DNA is undamaged; its long length (48,502bp) means the band is smeared from the top, but the clear lower edge shows all the DNA in the band to be the same molecular weight. Lanes 2 and 3 show λ DNA treated with low temperature plasma for 5 seconds and 30 seconds respectively, this means the DNA is subjected to reactive oxygen and nitrogen species; the shorter, damaged λ DNA ran further from the well and shows smearing on the lower edge, indicating DNA of multiple lengths is present.

Agarose gel electrophoresis allows the visualisation of DNA damage from *in vitro* experiments, or from a cell colony from which the DNA has been collected. This has the advantage that all the DNA starts in the same place in a small well before electrophoresis, so theoretically the sizes of the different fragments can be found based on how far they have run, but it cannot yield information about where within the nucleus of a single cell the damage occurred. For information on the DNA damage within a single cell the comet assay is used.

The comet assay (first developed by Östling & Johansson (Ostling and Johanson 1984) in 1984 and later modified by Singh et al. in 1988 (Singh et al. 1988) is often used for quantifying DNA damage in single cells. Cells are embedded in low-melting point agarose, cast onto microscope slides, lysed, electrophoresed and stained. As with gel electrophoresis, the DNA then runs from its starting position a distance that depends on its weight and structure, and the resulting shape of DNA gives the technique its name; "comet". The difficulties with this method come in evaluation- the DNA's initial position in the nucleus prior to electrophoresis must be determined, and in samples that have high levels of DNA damage this can be non-trivial and makes analysis subjective. Additionally, cells and their resulting comets can overlap making analysis difficult or impossible. The major advantage of this assay at the time of development was that compared to other methods it did not require radioactive labelling.

Software such as OpenComet (Gyori et al. 2014), a plug-in for ImageJ (http://imagej.nih.gov/ij/ ) can reduce the subjectiveness of manual analysis of the comet assay, are fast, and in this instance, free and open source (which allows bespoke modification).

DNA damage *in vitro* obviously has differences to DNA damage *in vivo*; for one, buffer or free-radical scavenging properties are changed. This, combined with slight differences in the method used to inflict damage make comparisons of the two difficult. For example, with low temperature plasma treatments the plasma source itself, distance from the plasma source and detection method used can vary (For example, contrast A M Hirst et al., 2015 and O'Connell et al., 2011). For low temperature plasma treatments at least, dose times to produce DNA damage generally fall in the range of seconds for low damage levels up to around 10 minutes for high damage levels, both *in vitro* and *in vivo*, indicating that DNA damage occurs on a similar timescale in both methods.

2.2 Single molecule qPCR

A technique to quantify RNA transcription is single molecule qPCR. This uses PCR in a single cell to quantitatively analyse transcription products: choosing a cell, and quantitative PCR can be achieved with reasonably high throughput in a microfluidic device (White et al. 2011). This technique is used to analyse the most abundant mRNA in a cell, but with further sensitivity and careful primer choice could plausibly be extended to examine damage to the DNA in the cell, but this is not a reality yet.

2.3 Immunofluorescence

Immunofluorescence labelling (Coons et al. 1942; Coons and Kaplan 1950) uses fluorescently labelled antibodies to optically detect antigens in tissue. This is a fluorescence technique giving data about single cells, but not quantitatively. Immunofluorescence labelling for visualising DNA damage from a plasma source (Hirst, Frame, Maitland, & O'Connell, 2014) gives you more spatial information than the comet assay, since the DNA is not moved from its original position. It does however produce difficulties due to the three dimensional nature of the cell, causing loss of information out of the focal plane and the potential for vertically overlapping signals being mistaken for single molecules.

2.4 Tunel staining

Tunel staining (Gavrieli et al. 1992), is a single cell fluorescence technique that fluorescently labels the 3' OH of damaged DNA produced by programmed cell death. The method requires staining followed by imaging, and is not without problems (Kraupp et al. 1995). However, the 3' OH group is not present in drug-induced DNA damage and therefore, although this is a fluorescent method, it will not be further discussed here.

3 Putative Fluorescence Single Molecule DNA damage Assays

The assays described above based on gel electrophoresis produce a measure of the size of the DNA fragments, but no information on where along the strand or in the chromosome the DNA damage occurred, or whether one region was a particular target.  The method of immunofluorescence suffers from uncertainty due to the three dimensional nature of the cell, and single molecule qPCR for DNA damage has not yet been achieved. Single molecule assays are preferable for applications where the damage caused is expected to be strongly dependent on distance from an application of a DNA damage source, for example cells at different distances from the point of application of a cancer treatment. In this case, ensemble measurements are insufficient as they do not give the required level of detail (Wollman et al. 2015).  In the following a putative fluorescence single molecule DNA damage quantification experiment will be described. Such an experiment requires a suitable substrate and a technique.

3.1 Substrate: DNA Origami

Watson-Crick base pairing of DNA can be used to self-assemble a long, single-stranded DNA strand ('scaffold') with complementary short strands ('staples') (Rothemund 2006): For example, a staple which binds to, say, sixteen bases in one location on the scaffold, and sixteen in another will act to bring those two regions together. By complementing the whole scaffold sequence with staples in this way, any shape can be designed.

Originally the correctly folded yield of DNA origami was very variable (in the Rothemund paper (Rothemund 2006) the 'rectangle' yield was 90% but the 'square' yield was 13%). Steps have been taken to increase this; some of these are empirical (see for example Martin & Dietz, 2012), but optimisation of the folding pathway (Dunn et al. 2015) gives the potential for more systematic improvement of yield, potentially making commercial production viable - this would be required for an object to be used as a standard DNA damage probe.

DNA origami might prove to be a good test object for DNA damage due to its 2D nature and well defined shape, thus circumventing the problems of immunofluorescence. Gel electrophoresis has been used as a method of analysing the folding quality of origami in empirical folding optimisation experiments (See for example Schmied et al., 2014). Double strand breaks may manifest in similar ways to incompletely folded origami. If so, this would mean DNA origami could be used to test a fluorescence single molecule DNA damage assay, as results could be compared to the results of the established method of gel electrophoresis.

3.2 Technique: Single molecule fluorescence microscopy

Fluorescence microscopy is a relatively non-invasive single molecule method; small fluorescent markers can be attached to molecules of interest *in vitro* or *in vivo*. What can be seen in conventional fluorescence microscopy is restricted: The Abbe diffraction limit gives a theoretical limit on the resolution of a light microscope of around two hundred nanometres, caused by diffraction through the rear aperture of the objective (for a discussion of this see Wollman et al. 2015). This is a problem, since many biological structures of interest are smaller than this, being just a few nanometres in size, so simply cannot be seen by looking down a conventional light microscope. Super-resolution fluorescence microscopy can be achieved by many methods (e.g. STED (Hell and Wichmann 1994), STORM (Rust et al. 2006),PALM (Betzig et al. 2006; Hess et al. 2006). What all of these methods have in common is they produce images of the point spread function of a single fluorescent molecule, (for example by spatially separating fluorescent molecules by using a low density, or by temporally separating the emission of densely packed molecules). The shape of the emission profile of a single molecule is known to be an Airy function, and by numerically fitting the centre of the distribution the location of the molecule can be found to a precision of nanometres, allowing the observation of interactions or the quantification of the number of molecules present.

A fluorescence microscopy, single-molecule tool for detecting DNA damage that uses a two dimensional DNA test object would be attractive for quantifying *in vitro* DNA damage as it removes the need to measure migration of DNA as in gel electrophoresis, and removes the issues inherent in the 3D nature of immunofluorescence. Intercalating dyes such as YOYO-1 increase in brightness intensity on binding to double stranded DNA by 100-1000 times their typical intensity in solution (Flors 2013). By using an intercalating dye at a concentration to fully saturate the origami tile, the fluorescence intensity of an undamaged tile could be measured. Single-molecule fluorescence microscopy is an increasingly quantitative tool enabling information to be extracted concerning molecular stoichiometry, mobility and co-localization (Leake 2013; Robson et al. 2013; Llorente-Garcia et al. 2014; Leake 2014). By finding the characteristic intensity of a single dye and combining this with measurements of the intensity of dye saturated damaged DNA origami, a measure of the number of double strand breaks could be made.

A microscope for such work is currently under development, incorporating a total internal reflection fluorescence (TIRF) capability to further reduce background from out of focus dye molecules. Figure 4 shows the setup. The excitation light can pass through either of two paths which contain lenses on flip mounts for different excitation modes. One path can be used as a FRAP path or an epifluorescence path, whilst the other can be operated in widefield or Slimfield/narrowfield (Plank et al. 2009; Wollman and Leake 2015) , in TIRF or epifluorescence modes. For imaging of DNA origami illumination is via the narrowfield TIRF mode. Lens L6 is mounted on a translation mount. By positioning L6 the focal length of L7 from L7, L6 lies in a conjugate plane to the focal plane. This means that angle of the beam for TIRF can be adjusted with no lateral deflection of the laser beam in the sample plane. This is beneficial since the physical size of the EMCCD (electron multiplying charge-coupled device) sensor on an Andor Ixon Ultra 897 is ~8.2mm x 8.2mm and large shifts would require realignment, making switching between illumination modes non-trivial.

This microscope combines fast high magnification (200X) imaging with bespoke TIRF illumination in a colour channel of choice. This should aid the acquisition of high signal-to-noise ratio images for precise quantification of the number of single dye molecules present on a DNA origami substrate to enable quantitative measures of DNA damage.

A fluorescence microscopy tool still has limitations – the production of free radicals such as molecular oxygen during laser illumination is a well-known factor (Dave et al. 2009). These free radicals cause damage to DNA, for example in an assay to look at stretched lambda DNA without oxygen scavenging (Miller et al. 2015) DNA strands spontaneously undergo snapping behaviour presumably due to radiative damage. Clearly a tool to image DNA damage should not itself inflict DNA damage by the same method. Luckily, the effects of illumination can be mitigated somewhat with oxygen scavenging systems; detailed studies have been conducted into how effective these are (See for example Dave et al. 2009).

4 Conclusions

There exist bacterial pathogens that act through DNA damage. Whilst several methods exist to quantify damage, none use the exact same target for each treatment. A DNA origami fluorescence microscopy assay would aid reproducibility and has advantages over techniques such as immunofluorescence since the entire DNA tile can be in focus in one image. A fluorescence microscopy assay to image DNA damage on single origami tiles is under development.

5. Acknowledgements

This work was funded by with the assistance of the Biological Physical Sciences Institute (BPSI) and the Royal Society (MCL). Thanks to Katherine Dunn, Adam Hirst and Deborah O'Connell for discussions and technical assistance on DNA origami and plasma treatments respectively.

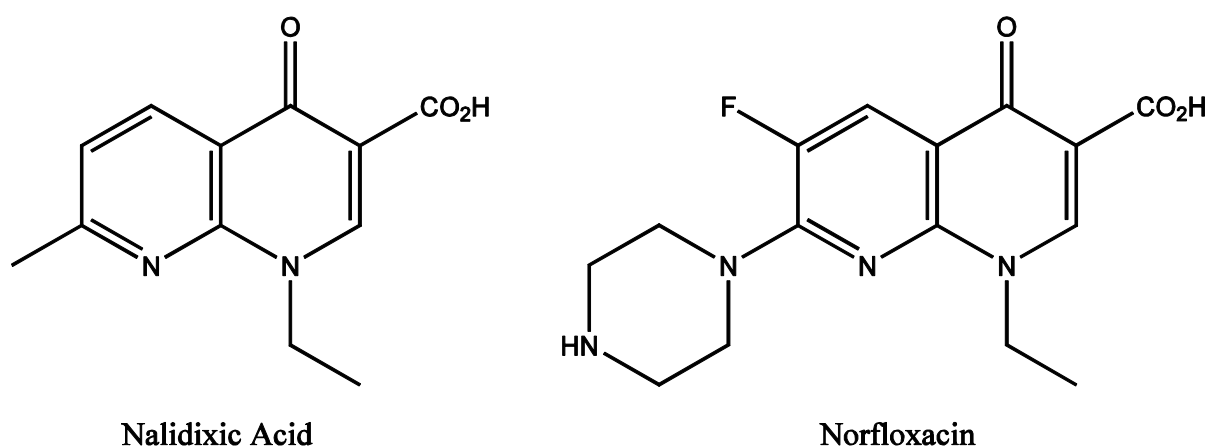

Nalidixic Acid                    Norfloxacin

Figure 1: The chemical structures of Nalidixic acid and Norfloxacin. Norfloxacin is an example of a fluoroquinolone.

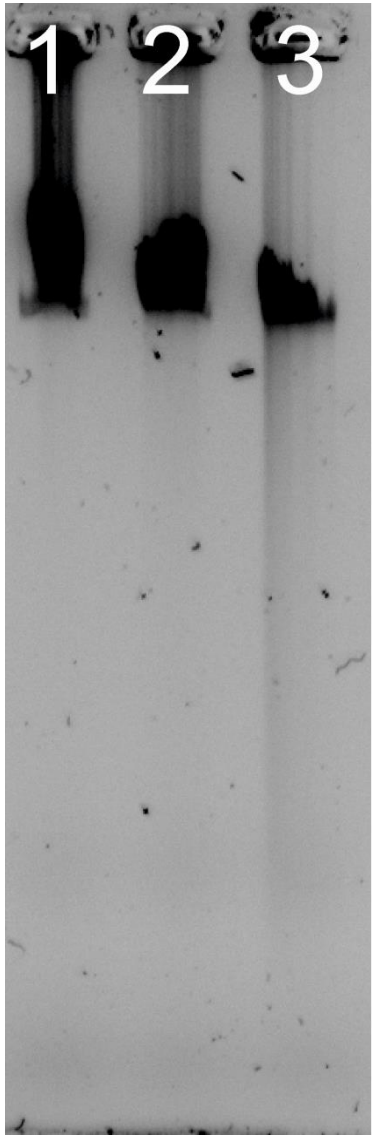

Figure 2: 1% agarose gel showing damage to λ DNA. Lane 1: Undamaged λ DNA, Lane 2: λ DNA treated with plasma for 5 seconds. Lane 3: λ DNA treated with plasma for 30 seconds.

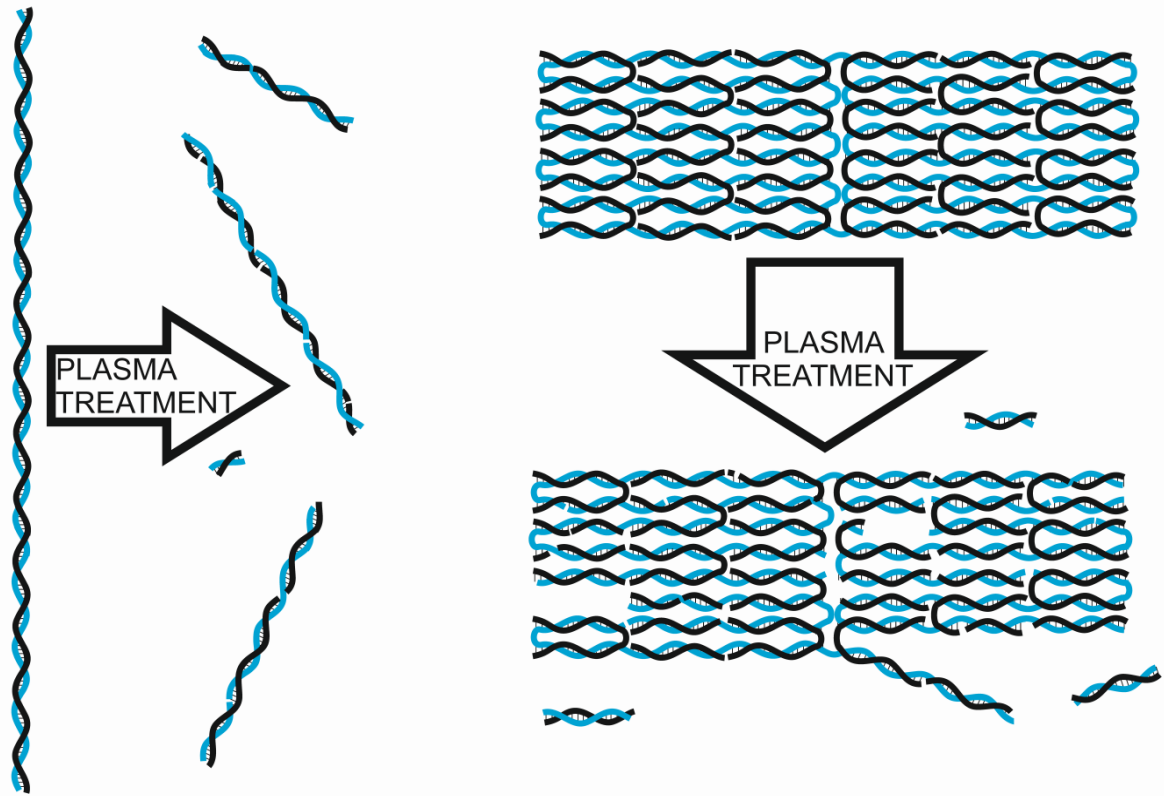

Figure 3: Examples of linear DNA and DNA origami substrates for a fluorescence damage assay, and a possible interpretation of the effects of low-temperature atmospheric plasma on each one (single and double strand breaks). The broken nature of the linear DNA lends itself to gel electrophoresis quantification as the different length pieces of DNA run to different positions. Damage to the origami tile could be quantified by total fluorescence intensity when imaged with an intercalating dye.

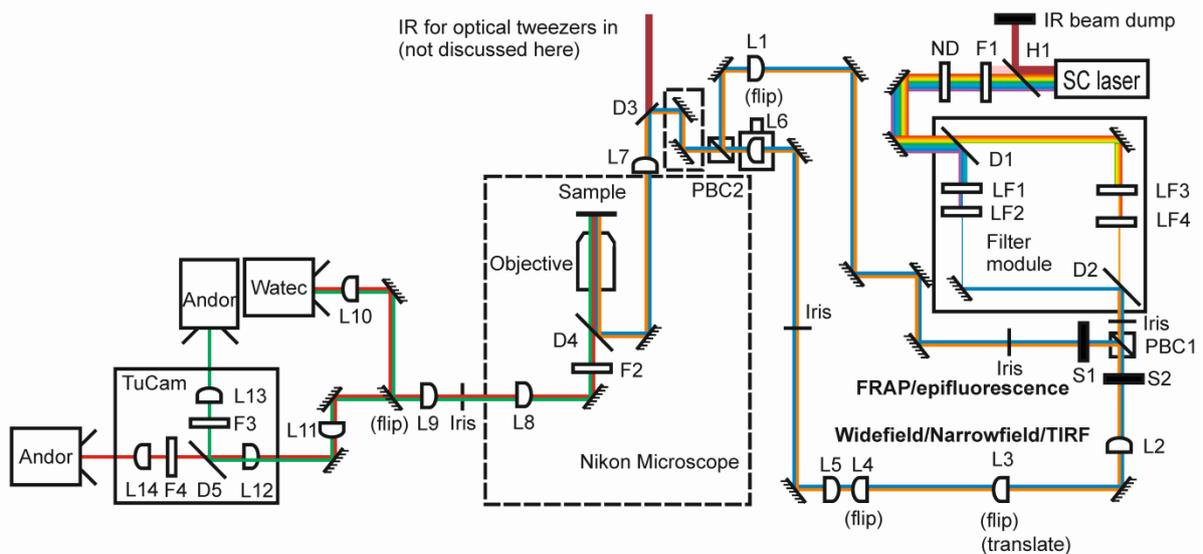

Figure 4: The bespoke fluorescence microscope designed for DNA damage assays. Lens numbers as described in the text. The infrared (IR) beam is part of ongoing development to add optical tweezers to the microscope (Zhou et al. 2015).